\documentstyle[12pt]{article}

\newcommand{\sect}[1]{\setcounter{equation}{0} \section{#1}}
\newcommand{\be}{\begin{equation}}
\newcommand{\ee}{\end{equation}}
\newcommand{\g}{{\tilde{g}}}
\newcommand{\p}{{\hat{\phi}}}
\begin{document}
\title{Centrifugal force in Kerr geometry}
\author{Sai Iyer and A R Prasanna\\
Physical Research Laboratory\\
Ahmedabad 380009\\
INDIA}
\date{}
\maketitle
\begin{abstract}
We have obtained the correct expression for the centrifugal
force acting on a particle at the equatorial circumference of a
rotating body in the locally non-rotating frame of the Kerr
geometry. Using this expression for the equilibrium of an
element on the surface of a slowly rotating Maclaurin spheroid,
we obtain the expression for the ellipticity (as discussed
earlier by Abramowicz and Miller) and determine the radius at
which the ellipticity is maximum.
\end{abstract}
\sect{Introduction}

The reversal of centrifugal force at the photon circular orbit of the
Schwarzschild geometry as shown by Abramowicz and Prasanna~\cite{AP}~(AP)
 has had several
interesting consequences and could in principle account for the
maximum of the ellipticity of a collapsing Maclaurin spheroid
preserving mass and angular momentum as demonstrated by
Abramowicz and Miller~\cite{AM}~(AM).

In order to discuss the case of a rotating body, it is indeed
necessary to use the Kerr geometry and consider the 3+1
splitting and identify the forces.  Prasanna and Chakrabarti~\cite{PC}
had considered the study of angular momentum coupling
using the optical reference geometry~\cite{ACL} in Kerr
spacetime and had obtained the separation of the total
four-force acting on a test particle in terms of gravitational,
centrifugal and Coriolis forces. However, while considering the
conformal splitting, they had used the Boyer-Lindquist form of
the Kerr metric which has the restriction of ergosurface being
the static limit surface. Hence, in that splitting, the
discussion of the behaviour of forces can be made only beyond
the ergosurface ($r=2$ in the equatorial plane) and not from the
event horizon onwards as in the Schwarzschild case. This indeed
can be rectified by using the locally non-rotating frame (LNRF), which,
in fact, is the most suitable one for discussing the dynamics of
rotating configurations in general relativity.

In the following we start with the Kerr geometry in
LNRF and then, using the conformal splitting, obtain the
expression for the centrifugal force. This  is then used in the
case of a slowly rotating Maclaurin spheroid to obtain the
expression for its ellipticity. We follow the same notation as
in the earlier papers referred to.
\sect{Formalism}

The Kerr metric as expressed in LNRF is given by
\be
ds^2 = -\frac{\Sigma\Delta}{B}dt^2 + \frac{\Sigma}{\Delta}dr^2
       +\Sigma d\theta^2 + \frac{B}{\Sigma}\sin^2\theta d{\hat\phi}^2,
\ee
where
\begin{equation}
\begin{array}{l@{\mbox{\hspace{0.5in}}}l}
d\hat\phi = d\phi - \omega dt\;, & \Sigma = r^2+a^2\cos^2\theta,\\
\Delta = r^2-2mr+a^2\;, & B = {(r^2+a^2)}^2 - \Delta a^2\sin^2\theta
\end{array}
\end{equation}
Considering the 3+1 conformal splitting as introduced by Abramowicz,
Carter and Lasota~\cite{ACL},
\begin{eqnarray}
ds^2 & = & \Phi\left[-dt^2 + d{\tilde{l}}^2\right],\nonumber\\
d{\tilde{l}}^2 & = & {\tilde{g}}_{\alpha\beta}dx^\alpha dx^\beta,
\end{eqnarray}
we have
\be
\begin{array}{l}
\Phi=\Sigma\Delta/B,\\
\g_{rr}=B/{\Delta^2},\quad\g_{\theta\theta}=B/\Delta,
\quad\g_{\p\p}=B^2/(\Delta\Sigma^2).
\end{array}
\end{equation}
Restricting the functions to the $\theta=\pi/2$ hyperplane we have
\begin{eqnarray}
\Phi & = & r\left(r^2-2mr+a^2\right)/\left(r^3+a^2r+2ma^2\right),\nonumber\\
\g_{rr} & = & \g_{\theta\theta}/\Delta =
\left(r^4+a^2r^2+2ma^2r\right)/{\left(r^2-2mr+a^2\right)}^2, \nonumber\\
 \g_{\p\p} & = &
{\left(r^3+a^2r+2ma^2\right)}^2/\left[r^2\left(r^2-2mr+a^2\right)\right].
\end{eqnarray}

Using the same notation as in AP, it can be easily verified
that the spatial 3-momentum~$p^i$  defined through $p^i = \Phi P^i$,
where $P^i$ is the 3-momentum in the four-space, may be
used to evaluate the centrifugal acceleration.  For a particle
in a circular orbit the invariant speed~$\tilde{v}$ in the projected
manifold is given by
\be
\label{vtilde}
{\tilde{v}}^2 = \Phi g_{ij} u^i u^j,
\ee
where $u^i$ is the four velocity of the particle.  Using the
usual definition of the four velocity in terms of Killing
vectors, $u^\alpha = A\left(\eta^\alpha+\hat{\Omega}\xi^\alpha\right)$,
where $A$ is the redshift factor, one gets $u^i = A\hat{\Omega}\xi^i$
and then
\be
{\tilde{v}}^2 =
\frac{{\hat{\Omega}}^2 r^2}{1-{\hat{\Omega}}^2{\tilde{r}}^2}
= \frac{\Phi\hat{\Omega}^2\tilde{r}^2}{1-\hat{\Omega}^2\tilde{r}^2}
= \frac{\hat{\Omega}^2 \g_{\p\p} g_{tt}}{1-\hat{\Omega}^2 \g_{\p\p}}
\ee

The geodesic curvature radius~$\cal{R}$ defined through~\cite{AP}
\be
{\cal R} = \tilde{r}/{\left|\g^{ij}\left(\tilde{\nabla}_i\tilde{r}\right)
\left(\tilde{\nabla}_j\tilde{r}\right)\right|}^{1/2}
\ee
projected onto the instantaneously corotating frame with the
spatial triad
\be
e^r_{(r)}=(\g^{rr})^{1/2},\quad e^\theta_{(\theta)}=(\g^{\theta\theta})^{1/2},
\quad e^\p_{(\p)}=(\g^{\p\p})^{1/2}
\ee
is given by
\begin{eqnarray}
\cal{R} & = & \tilde{r}/\left(\partial_r\tilde{r}\right)
          = 2\g_{\p\p}/\left(\partial_r\g_{\p\p}\right) \nonumber\\
& = & \frac{r\Delta\left(r^3+a^2r+2ma^2\right)}
{\left(r^5-3mr^4+a^2r^3-3ma^2r^2+6m^2a^2r-2ma^4\right)}.
\end{eqnarray}
If $L$ denotes the angular momentum as measured by the stationary
observer (instantaneously at rest), one has
\be
L=\xi^a u_a = A\hat{\Omega}(\xi\xi) = A\hat{\Omega}g_{\p\p}
\ee
and thus from equation (\ref{vtilde}) and the definition of $u^i$,
\be
L^2 = \tilde{v}^2 \g_{\p\p}.
\ee
Hence if we now consider the definition of centrifugal force as
given in AP, $C_f = m_0 \tilde{v}^2/\cal{R}$, we get
\be
\label{cfgen}
C_f = \frac{m_0 L^2}{\g_{\p\p}}\frac{\partial_r\g_{\p\p}}{2\g_{\p\p}}
=\frac{m_0 L^2}{2}\frac{\partial_r\g_{\p\p}}{(\g_{\p\p})^2},
\ee
\be
\label{cf}
C_f = \frac{L^2 r \left(r^5-3mr^4+a^2r^3-3ma^2r^2+6m^2a^2r-2ma^4\right)}
       {{\left(r^3+a^2r+2ma^2\right)}^3}.
\ee
It is interesting to note that the final expression for the
centrifugal force (eq.~\ref{cfgen}) just turns out to be exactly
\be
{\left(p^\p\right)}^2 \partial_r \g_{\p\p}/2,
\ee
as was obtained for the static spacetime.
The parameters~$L$ and~$\hat{\Omega}$, the angular velocity
measured in the global inertial frame, are related through the
expression
\be
L=\frac{\hat{\Omega}\left(r^2+a^2+2ma^2r\right)^{3/2}}
{\left[\left(r^2-2mr+a^2\right)-\hat{\Omega}^2
\left(r^2+a^2+2ma^2/r\right)^2\right]^{1/2}}.
\ee

It is indeed clear from the expression~(\ref{cf}) that for all real
values of $L$, the {\it centrifugal force reverses sign} at the
zeros of the function
\be
f(r) = r^5-3mr^4+a^2r^3-3ma^2r^2+6m^2a^2r-2ma^4.
\ee
Using Sturm's theorem one can ascertain that there are three
real roots of the equation $f(r)=0$ for $0\le r<\infty$.  Of
these three real roots, two lie between $r=0$ and the event
horizon and are thus of no consequence to any outside observer.  The
third root lies between  $r= 2m$ and $r = 3m$, i.e., the ergosurface and the
surface where centrifugal reversal occurs for non-rotating
objects ($a=0$).  Table~(1) enlists the values of $R$ ($=r/m$) for which
$f(R)=0$ for
different values of $\alpha$ ($=a/m$).
\begin{table}
\caption{Location of the zero of $C_f$ for different $\alpha$.}
\begin{center}
\begin{tabular}{||cc|cc||}
\hline\hline
$\alpha$&$R$&$\alpha$&$R$\\ \hline
0.0&3.0000&0.6&2.9202\\
0.1&2.9978&0.7&2.8916\\
0.2&2.9911&0.8&2.8590\\
0.3&2.9800&0.9&2.8226\\
0.4&2.9645&1.0&2.7830\\
0.5&2.9445&&\\ \hline\hline
\end{tabular}
\end{center}
\end{table}
As $\alpha$ increases from 0 to 1 the surface
of reversal moves inwards from  $R= 3$.
Fig.~(1) shows the behaviour of the centrifugal force for
non-zero values of $\alpha$, depicting the inward movement of
the zero and the maximum as well as the flattening of the arm
beyond the maximum.

\sect{Slowly rotating configurations}

For slowly rotating Maclaurin spheroids the balance of forces
on a surface element in the equatorial plane is expressed
through the equation~\cite{AM,C}
\be
\label{mfbal}
\mbox{(Centrifugal force)} = \mbox{(Gravitational force) }f(e),
\ee
with $e$ denoting
the eccentricity of the spheroid, expressed through the standard
relation, and $f(e)$ is given by
\be
f(e)=\frac{9(1-e^2)}{2e^2} \left[\frac{\sin^{-1}e}{\sqrt{1-e^2}}
\left(1-\frac{2e^2}{3}\right) -e\right].
\ee
The ellipticity~$\epsilon$ is expressed in terms of $e$ by the relation (AM)
\be
\epsilon = \left[1-(1-e^2)^{1/2}\right]/(1-e^2)^{1/6}.
\ee
Following the same approximation as in \cite{AM}, we take for slow
rotation
\be
\epsilon=e^2/2, \quad f(\epsilon)=4\epsilon/5.
\ee
Further the angular momentum~$J$ turns out to be
\be
\label{angmom}
J = \frac{2}{5}(1-e^2)^{-1/3}L \approx 2L/5.
\ee
Using equations~(\ref{angmom}) and~(\ref{cf}) in equation~(\ref{mfbal})
we get, without any approximation in terms of $\hat{\Omega}$ or $a$,
\be
\epsilon=\frac{125}{16}J^2
\frac{R^3(R^5-3R^4+\alpha^2R^3-3\alpha^2R^2+6\alpha^2R-2\alpha^4)}
{(R^3+\alpha^2R+2\alpha^2)^3},
\ee
with $R = r/m$ and $\alpha=a/m$.  Thus  $\bar{\epsilon}=\epsilon/J^2$
expressed in terms of $R$ is
\be
\bar{\epsilon}(R)=\frac{125}{16}
\frac{R^3(R^5-3R^4+\alpha^2R^3-3\alpha^2R^2+6\alpha^2R-2\alpha^4)}
{(R^3+\alpha^2R+2\alpha^2)^3}.
\ee
Fig.~(2) shows the behaviour of $\bar{\epsilon}$ as a function of $R$
and it may be seen that the maximum in $\bar{\epsilon}$, which appears
at $R = 6$ for $\alpha = 0$,
now moves slightly outwards for $\alpha \neq 0$.
Thus one finds that taking
into account the potential of the Kerr metric for the surface
element of the rotating fluid configuration, the maximum of
ellipticity occurs earlier during the collapse (i.e., for larger
radii) than what was found using the Schwarzschild potential.
It may also be seen directly that it is not necessary to make
any approximation in the expression with respect to $\hat{\Omega}$ or $a$ to
find the ellipticity in units of $J^2$.  This is, in fact, true
for the Schwarzschild case also. (In the paper of AM they had
neglected the terms in $\Omega^2$ for slow rotation, which in fact is not
necessary).

Table~(2) lists the location of the occurence of
$\bar{\epsilon}_{\mbox{\scriptsize max}}$
for different $\alpha$  and also the values of
$\bar{\epsilon}_{\mbox{\scriptsize max}}$ and
the centrifugal force at that location.
\begin{table}
\caption{Location of the maximum of $\bar{\epsilon}$ for different
$\alpha$ and the values of $\bar{\epsilon}$ and $C_f$ there.}
\begin{center}
\begin{tabular}{||cccc||}
\hline\hline
$\alpha$&$R$&$\bar{\epsilon}_{\mbox{\scriptsize max}}$&$C_f$\\ \hline
0.0&6.0000&0.6510&0.0145\\
0.1&6.0044&0.6506&0.0144\\
0.2&6.0178&0.6491&0.0143\\
0.3&6.0401&0.6467&0.0142\\
0.4&6.0713&0.6435&0.0140\\
0.5&6.1116&0.6393&0.0137\\ \hline\hline
\end{tabular}
\end{center}
\end{table}
As  $\bar{\epsilon}$
expresses the ratio of $C_f$ to the gravitational force and
since $C_f$ flattens after the maximum for increasing $\alpha$
(see Fig.~1),
the location of $\bar{\epsilon}_{\mbox{\scriptsize max}}$ shifts
outward for increasing $\alpha$.
\sect{Concluding remarks}

The study of equilibrium configurations for rotating fluid masses, a topic
which has been analysed thoroughly in the Newtonian framework,
is still to find a completely satisfactory analysis in general
relativity, due to the lack of exact solutions.  However, there have
been a number of investigations to study the effects of general
relativity on the dynamics of rotating fluids bearing out the
necessity to improve the Newtonian results by including GR effects.
The Kerr geometry, which is an exact representation of the field
outside the rotating body, should be the correct framework to be
used for studying the effects of rotation, particularly through
the Locally Non-Rotating Frames introduced by Bardeen~\cite{B}.  In most of
the studies carried out earlier, when one used the
Boyer-Lindquist coordinates one could not separate out the
inertial frame dragging effects, leading to ambiguous
interpretations.  Thus in the analysis made above using LNRF we
have first obtained the correct expression for the centrifugal
force acting on the surface element (in the equatorial plane) of
a rotating mass and then studied the behaviour of the
ellipticity as was discussed by Chandrasekhar, Miller and
Abramowicz~\cite{AM,CM,M}.

The main result that the inclusion of the Kerr potential brings
is that the ellipticity maximum occurs at a radius slightly
larger than in the case of the Schwarzschild potential.  This can
certainly have some consequence in the dynamics of rotating
compact stars.  In fact, it is known that most of the
calculations of neutron star models assume that there is a
centrifugal barrier and thus confine the rotation speed to
certain limits, whereas if one considers the effect of general
relativity through centrifugal force reversal there can be an effect both
on the shape and size of the star different from what is assumed.
As pointed out by AM, it is still necessary to include the
effect of pressure gradient forces in a compatible way along
with the correct form of centrifugal force as given in equation~(\ref{cf}) to
get a complete picture of the shape and size of a rotating star.

\end{document}